\documentclass[10pt,conference]{IEEEtran}
\IEEEoverridecommandlockouts
\usepackage{cite}
\usepackage{amsmath,amssymb,amsfonts}
\usepackage{algorithmic}
\usepackage{graphicx}
\usepackage{textcomp}
\usepackage{dblfloatfix}
\usepackage{xcolor}
\def\BibTeX{{\rm B\kern-.05em{\sc i\kern-.025em b}\kern-.08em
    T\kern-.1667em\lower.7ex\hbox{E}\kern-.125emX}}
\begin{document}

\title{Thunder CTF: Learning Cloud Security on a Dime
}

\author{\IEEEauthorblockN{Nicholas Springer}
\IEEEauthorblockA{
\textit{Portland State University}\\
nicholas.m.springer@gmail.com}
\and
\IEEEauthorblockN{Wu-chang Feng}
\IEEEauthorblockA{
\textit{Portland State University}\\
wuchang@pdx.edu}
}

\maketitle

\begin{abstract}
Organizations have rapidly shifted infrastructure and applications over to public cloud computing services such as AWS (Amazon Web Services), Google Cloud Platform, and Azure. Unfortunately, such services have security models that are substantially different and more complex than traditional enterprise security models. As a result, misconfiguration errors in cloud deployments have led to dozens of well-publicized breaches. This paper describes Thunder CTF, a scaffolded, scenario-based CTF (Capture-the-Flag) for helping students learn about and practice cloud security skills. Thunder CTF is easily deployed at minimal cost and is highly extensible to allow for crowd-sourced development of new levels as security issues evolve in the cloud.
\end{abstract}


\section{Introduction}
An overwhelming percentage of companies are leveraging public cloud services for their computing infrastructure, finding its economies of scale hard to pass up \cite{451cloud}. As companies move infrastructure and applications to public cloud platforms such as AWS, Google Cloud Platform, and Azure, it is becoming increasingly important for practitioners to be aware of security issues that may lead to compromise. More than 2/3 of all enterprises list security as the most significant concern for moving to the cloud \cite{columbus18cloud}. Such concern is not unfounded, as the cloud presents its users with a new set of operational security configurations that can be difficult to comprehend and set up properly. As a result, breaches ranging from the discovery of unprotected cloud resources such as storage buckets \cite{osullivan19rnc} and search indexes \cite{diachenko19document}, to more sophisticated application compromises \cite{walikar19ssrf} have led to the exposure of millions of financial and voter records. The shift to the cloud has added significant complexity to the practice of security. While legacy infrastructure must deal with machines, operating systems, routers, firewalls, software patching of vendor-supported software, as well as username and password management, cloud infrastructure must handle all of those issues potentially, while also dealing with role-based access control, zero-trust networks, API security, account access keys, authentication tokens, and federated identity providers. Additionally, cloud infrastructure security is moving increasingly into the hands of software development teams as well as a vast open-source software supply chain that no one in particular is in charge of, making things even more difficult to secure.

Learning cloud security can be a daunting task with the broad range of topics to learn and skills to practice. One popular way is through Capture-the-Flag (CTF) exercises, which challenge players to find security flaws and execute exploits on intentionally vulnerable systems. CTF exercises have been successfully used as both a vehicle for experienced practitioners to sharpen their skills and for beginners to develop them. Unfortunately, with cloud security in a formative state, there are few CTFs that focus on providing guidance for beginners to learn about cloud security. Additionally, existing CTFs focus exclusively on securing AWS deployments \cite{cloudgoat, flaws, flaws2, serverlessgoat}. To address this, this paper describes Thunder CTF, a CTF and framework for helping both experienced and novice practitioners learn about securing projects on Google Cloud Platform \cite{thunder-ctf}.

Section~\ref{sec2} describes the overall design of the CTF including its initial levels and its framework for supporting extensibility. Section~\ref{sec3} describes the results of an initial deployment in an advanced elective course in our program. Finally, Section~\ref{sec4} concludes and describes future work.

\section{Curriculum and CTF}
\label{sec2}
Thunder CTF is designed with several overall goals. These goals include 1) modeling exercises on actual compromises and commonly found problems in existing cloud deployments, 2) scaffolding exercises to support differentiated instruction for the benefit of both novices and experienced practitioners, 3) creating an extensible framework that allows developers to add, remove, and customize levels based on current vectors of exploitation, and 4) making the CTF easily deployable to minimize the friction of setting it up and the cost to run it.

\subsection{Scenario-based}
Rather than focusing on a single concept or skill, Thunder CTF levels are scenario-based and tied to actual compromises in order to better motivate the skills being practiced and to provide students a more engaging, realistic roleplaying experience. Scenarios are explicitly tied back to the real breaches that inspired them via a write-up at the end of each individual level, and scenarios often mirror the tactics and techniques enumerated in MITRE’s ATTACK matrix \cite{mitre-attack}: a framework for classifying typical adversarial behavior (e.g. initial access, persistence, privilege escalation, defence evasion, credential access, discovery, collection, exfiltration, and impact). This allows students to see a bit of the “forest” in which adversaries operate, rather than only looking at the "trees" of individual exploits.

Thunder CTF levels chain together multiple techniques and concepts in order to achieve an overall operational goal. Currently, the CTF consists of an initial set of 6 levels that cover important concepts in cloud security, such as open storage buckets, overprovisioned permissions, exfiltration of sensitive information via log files, security keys in source repositories, unsanitized error messages, access token compromise, backdoors via metadata, misconfigured IAM (Identity and Access Management) policies, IAM privilege escalation, exposed container images, and metadata credential compromise via server-side request forgery.

Table~\ref{tab1} displays the 6 different scenarios currently implemented in Thunder CTF. As the table shows, scenarios chain together security vulnerabilities in order to show students how they would function in a context of compromising a project.

\begin{figure}[b]
\centerline{\includegraphics[width=0.475\textwidth]{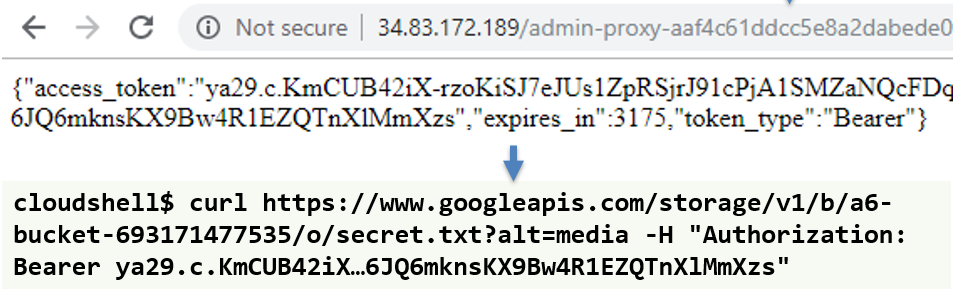}}
\caption{Capital One breach level}
\label{fig1}
\end{figure}

Fig.~\ref{fig1} shows a screenshot of a student solving a6container, a level that recreates a series of steps similar to those used in the Capital One breach \cite{walikar19ssrf}. As the figure shows, a vulnerable proxy service is found that the student can use to access the internal metadata service. The service exposes a session token that can be activated by the student to access storage buckets that the proxy has access to. Within one of these buckets, a file containing credit card information is then found.

\subsection{Scaffolded}
CTF levels often target a particular level of expertise. Unfortunately, students that find the levels too easy can become bored and stop playing, while those that find the levels too difficult can become frustrated and stop playing. To overcome this, Thunder CTF is designed to support differentiated instruction across users with disparate abilities. It features scaffolded levels with incrementally increasing difficulties as well as an extensive hint system. The hint system allows novices to make consistent progress through the CTF, while affording experienced players a challenging experience when the hints are ignored and the levels are completed in an open-ended manner. Such a setup also allows a student to revisit the CTF and replay its levels with less support from hints to solidify their skill level.

\begin{figure}[t]
\centerline{\includegraphics[width=0.475\textwidth]{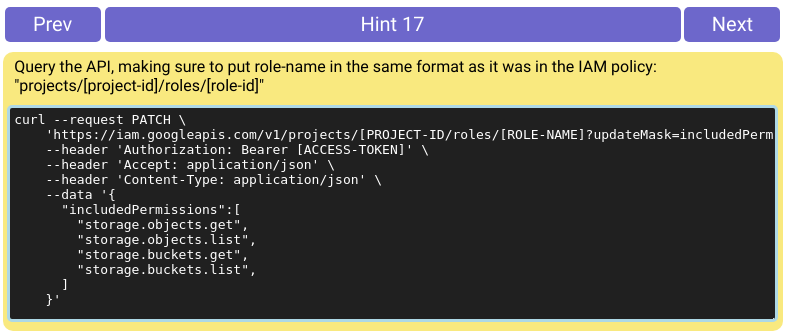}}
\caption{Hint system in Thunder CTF}
\label{fig2}
\end{figure}

Fig.~\ref{fig2} shows an example of a hint that is available for students to use on a particular level when they are stuck. Thunder CTF hints are sequentially accessed by users to guide them towards level completion. As the figure shows, explanations within the hints teach not only the concepts required to perform the next step, but also the syntax of the commands, and code required to run them. 

\subsection{Extensible and Modular}
Security issues are constantly evolving. Where vulnerabilities such as Unvalidated Redirects and Forwards and Cross-Site Request Forgery were in the OWASP Top 10 in 2013, they are now nowhere to be found. Similarly, while access token compromise via Metadata exploitation (an exploit used in level 6) may be possible now, by the time this paper is published, the addition of a required, custom HTTP request header may completely remove it as a vector of future compromise \cite{gcpmetadata, awsmetadata}. As a result of the rapidly changing landscape of security issues, Thunder CTF has been designed to be highly configurable and extensible, allowing levels to be quickly added and removed.

\begin{figure*}[!b]
\centerline{\includegraphics[width=0.8\textwidth]{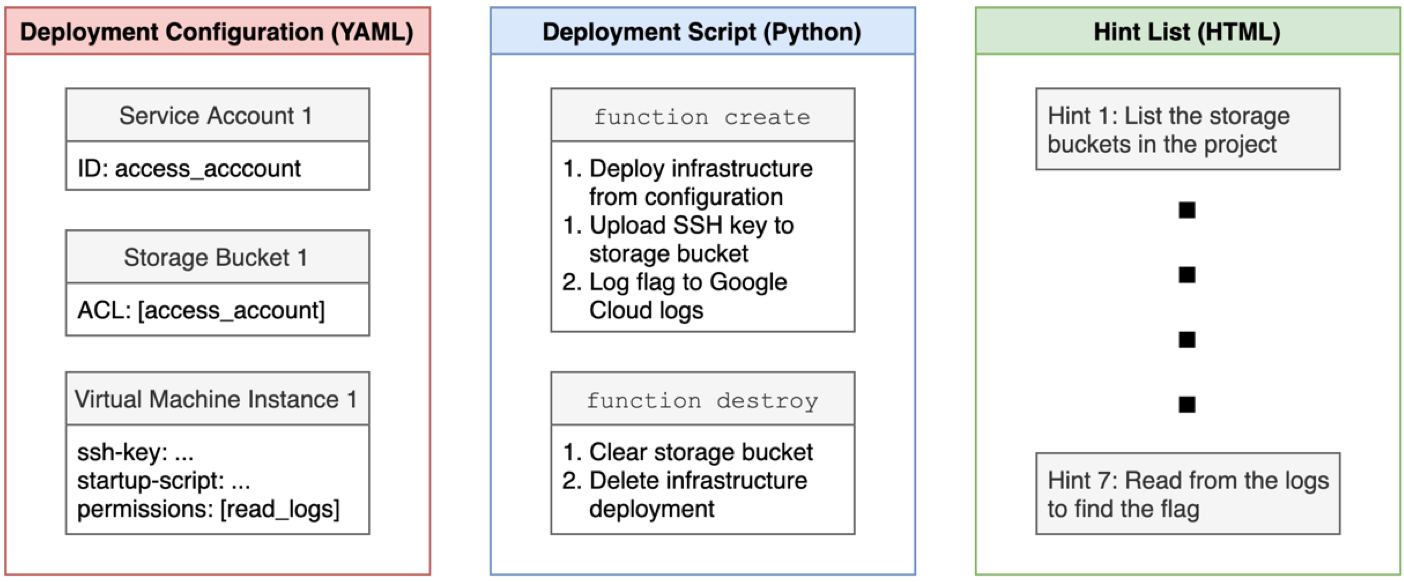}}
\caption{Outline of example level module}
\label{fig3}
\end{figure*}

Thunder CTF levels are modular and follow a specific structure, which reduces the process of creating new levels to simply filling in a template. Each level module contains as few as three files, which are displayed in Fig.~\ref{fig3}. These files include a deployment configuration that specifies the cloud resources needed for the level, a deployment script that encodes the logic for deploying and configuring infrastructure, and a hint list file containing the HTML content of each hint. A level development guide, which explains the process of writing each file of a level module, as well as a template level module, are provided for level creators in the Thunder CTF repository \cite{dev-guide}.

Initial infrastructure deployment is done using Deployment Manager, GCP’s infrastructure-as-code solution. The level creator provides a YAML deployment configuration that specifies the infrastructure required by the level, which the deployment script passes to the framework’s deployment interface. The deployment interface renders the configuration using a template engine, allowing level creators to customize configurations at runtime. Then, the deployment interface uses the Google Cloud Deployment Manager API to launch the level infrastructure. By abstracting away the specifics of dealing with the Deployment Manager API, the deployment interface allows level creators to focus on actual level-specific configurations. An outline of a possible deployment configuration is shown in the first column of Fig.~\ref{fig3}.

\begin{table*}[!t]
  \caption{Thunder CTF level scenarios}
  \begin{center}
  \begin{tabular}{|l|l|}
  \hline
  {\bf Scenario }& {\bf Walkthrough description } \\
  \hline
  {\tt a1openbucket} &
  \begin{minipage}[t]{0.6\textwidth}
    \textbullet{ List all storage buckets for a project using the command-line interface.}\\
    \textbullet{ Copy a secret file stored within a bucket left open}
  \end{minipage}
  \\
  \hline
  {\tt a2finance} & 
  \begin{minipage}[t]{0.6\textwidth}
    \textbullet{ Activate a service account credential.}\\
    \textbullet{ List permissions associated with a credential.}\\
    \textbullet{ Download the contents of an accessible bucket.}\\
    \textbullet{ Navigate a git repository that previously contained a sensitive {\tt ssh} key.}\\
    \textbullet{ Checkout repository version containing the initial commit of the key.}\\
    \textbullet{ List project's VM instances to find one containing an {\tt ssh} key in its Metadata.}\\
    \textbullet{ {\tt ssh} into the instance and use its role to access the project's logging service.}\\
    \textbullet{ Find credit-card numbers in log entries that have not been properly sanitized}
  \end{minipage}
  \\
  \hline
  {\tt a3password} &
  \begin{minipage}[t]{0.6\textwidth}
    \textbullet{ List all of the serverless functions for a project.}\\
    \textbullet{ Access an API endpoint for a function to discover it requires authentication.}\\
    \textbullet{ Obtain an identity token using your initial credentials and access API again.}\\
    \textbullet{ Reverse-engineer endpoint to discover it requires a password field.}\\
    \textbullet{ Use over-provisioned credential to download the function's source code from its REST API.}\\
    \textbullet{ Reverse-engineer function to expose a sensitive environment variable.}\\
    \textbullet{ List the function's Metadata to discover the secret and access the endpoint.} 
  \end{minipage}
  \\
  \hline
  {\tt a4error} &
  \begin{minipage}[t]{0.6\textwidth}
    \textbullet{ Obtain an identity token and use it to access a function via its URL.}\\
    \textbullet{ Inject unexpected input to trigger an error.}\\
    \textbullet{ Use credential to read the function's log entries.}\\
    \textbullet{ Find the function's credentials within the error log and list its permissions.}\\
    \textbullet{ Use the function's credentials to list the project's VM instances.}\\
    \textbullet{ Add an {\tt ssh} key to ethe Metadata of a VM instance.}\\
    \textbullet{ Log into the instance to access a secret file.} 
  \end{minipage}
  \\
  \hline
  {\tt a5power} &
  \begin{minipage}[t]{0.6\textwidth}
    \textbullet{ List credential permissions to find its ability to overwrite a serverless function.}\\
    \textbullet{ Overwrite function code to return credentials of the VM executing function.}\\
    \textbullet{ List VM's credential permissions to find its ability to view and edit IAM policies.}\\
    \textbullet{ View IAM policy via an API to discover its update role permission.}\\
    \textbullet{ Escalate privileges via an API to gain permissions to access storage buckets.}\\
    \textbullet{ Access secret file in storage bucket}
  \end{minipage}
  \\
  \hline
  {\tt a6container} &
  \begin{minipage}[t]{0.6\textwidth}
    \textbullet{ List the project's VM instances to discover a web server run via a container.}\\
    \textbullet{ Pull the container image and examine its code to find a hidden route implementing a proxy.}\\
    \textbullet{ Perform a server-side request forgery (SSRF) attack on the Metadata service for the VM to obtain its credentials.}\\
    \textbullet{ List VM's credentials to find its ability to access storage buckets.}\\
    \textbullet{ Access secret file in storage bucket}
  \end{minipage}\\
  \hline
  \end{tabular}
  \label{tab1}
  \end{center}
\end{table*} 

In many cases, after infrastructure is deployed, it must be configured further before the level is ready to be played. The deployment script of each level performs these configurations after the initial infrastructure deployment is complete. An outline of a possible deployment script is shown in the second column of Fig.~\ref{fig3}.

Infrastructure could be configured using Google Cloud REST APIs, but using the APIs directly complicates level development. To address this, the CTF framework provides helper functions that perform common infrastructure configurations, such as uploading files to storage buckets or modifying security policies. The abstraction of common API calls into helper functions allows level creators to focus on the overall configuration logic of their level, rather than dealing with the details of the underlying Google Cloud API calls. Documentation of the framework helper functions is provided on the Thunder CTF website \cite{framework-docs}.

An important aspect of the scaffolding in Thunder CTF is provided by the hint system. To build an attractive, functional hint system, a good deal of development work is required. To simplify the process for level creators, Thunder CTF provides a hint templating system that allows level creators to focus solely on producing the content of hints rather than the formatting. When creating a level, level creators can provide a hint list file that contains the content of each hint. Using the hint list as input, the framework generates a styled HTML document that displays the hints in a sequential slideshow. This simplification takes the burden of hint styling and delivery away from level creators, allowing them to focus on the content of the hints versus their presentation.

Finally, the framework of Thunder CTF provides extensibility both by supporting new level contributions within an existing CTF such as the 6 levels previously described, as well as by supporting the addition of entirely separate sequences of levels, effectively allowing content developers to design their own CTFs focused on new topics \cite{thunder-ctf}. For that, Thunder CTF has a namespace system that allows the same framework to implement a different set of levels with a correspondingly different site to serve the web content. Using this, we are currently implementing a separate CTF focused on web and application security issues that leverages the Thunder CTF framework for deployment in the cloud.

\subsection{Deployable}
Regardless of how well a CTF is designed, in order to be impactful, it must be widely used. Specifically, its code and content must be accessible, it must be easy to set up, and it must be either free to use or incur minimal cost. In addition, if being used in the context of a course, it must support levels that are polymorphic in nature to ensure each student has individually solved each level. Thunder CTF attempts to address each of these aspects.

\subsubsection{Freely available code and content}
The code itself is free (as in speech), and students download the entire code repository as part of running Thunder CTF. While the hint-system is included in the repository and can be used directly by students, for convenience, we provide a hosted site for accessing level instructions and hints \cite{thunder-ctf}. The repository is also structured to allow level contributions from the community in order to enable crowd-sourced development and deployment of new content.

\subsubsection{Frictionless setup}
The setup for Thunder CTF can be done within minutes. Students that are new to Google Cloud sign up for a free account with up to \$300 in credit \cite{gcpfree}. From a web browser, they then launch Cloud Shell, a GCP-based command line interface that is already set up for accessing cloud resources, and clone the Thunder CTF repository. From there, they run a single Python script to deploy the level they wish to play (e.g. python3 thunder.py create thunder/a1openbucket). The script interfaces with Google’s Cloud Deployment Manager API to programmatically set up all of the level’s resources on-demand. When the level is completed, the same script is used to bring down all of the resources that have been deployed (e.g. python3 thunder.py destroy). This simplistic interface makes it easy for students to deploy and navigate levels.

\subsubsection{Low cost}
Labs for cloud-based exercises often require a paid subscription to use \cite{infoseclearning}. Thunder CTF instead has been designed to be free (as in beer), with its resource consumption falling well within the free tier of Google Cloud. Specifically, each level utilizes virtual machines, cloud functions, containers, and API endpoints that fit within the “always-free” umbrella of Google Cloud. In addition, CTF levels make heavy use of serverless infrastructure that is billed only upon usage over a certain threshold. As a result, run-throughs of the CTF can cost as little as a dime, making the CTF a cost-effective way to learn and practice important cloud skills.

\subsubsection{Polymorphic levels}
Finally, for deployments in courses, it is important to ensure that each student performs individual work in completing the exercises, and that solutions to the levels can’t simply be shared amongst them. To address this, the CTF framework includes a function that generates unique secret flags, which students look for when playing each level. This is done by hashing the combination of a constant level seed and the GCP project identifier for the player’s cloud project. With this, a validation site can be easily built to check solutions on a per-user basis. Although the students have full access to CTF code, flags could potentially be reversed, but we believe it is easier to solve the levels than to reverse-engineer the flags from code.

\section{Evaluation}
\label{sec3}
The first use of Thunder CTF occurred in our Fall 2019 offering of Portland State University’s CS 430/530 Internet, Web, and Cloud Systems course with 48 students. The first half of the 10-week course covers key concepts in networking, operating systems, web development, and databases before transitioning to their use in cloud computing environments. At the beginning of the 6th week, a lecture on Google Cloud Identity and Access Management is given to students, followed by two weeks of labs that walk students through the basics of leveraging Cloud Storage, Cloud Datastore, App Engine, Cloud Run, Cloud Functions, and Kubernetes Engine to scale applications seamlessly. At the end of this sequence, the Thunder CTF levels were assigned to demonstrate how such services can be misconfigured in ways that expose them to attack. Students were given a due date for the exercises at the end of the 9th week, allowing them about a week to finish the levels.

To assess the effectiveness of the CTF, we surveyed students at the beginning of the 10th week. Table~\ref{tab2} lists the questions that were asked in the survey. Our goal was to measure how well the CTF helped students learn about cloud security issues and develop skills in navigating cloud systems. In addition, we were interested in measuring the utility of the hint system for supporting students as they completed the exercise.

\begin{table*}[!t]
  \caption{Survey questions for Thunder CTF in CS 430/530: Internet, Web, and Cloud Systems (Fall 2019)}
  \begin{center}
  \begin{tabular}{|l|}
  \hline
  {\bf Question }\\
  \hline
  Q1: Rate the CTF exercises for understanding security issues in the cloud.\\
  \hline
  Q2: Rate the CTF exercises for developing skills in navigating the cloud.\\
  \hline
  Q3: Rate the hint system as a mechanism for providing help as needed in solving CTF exercises.\\
  \hline
  \end{tabular}
  \label{tab2}
  \end{center}
\end{table*}

Of the 48 students in the class, 36 responded to the survey.  Table~\ref{tab3} shows the results. As the table shows, students felt that the lecture material and CTF exercises were both helpful for learning about security issues and for developing cloud skills, while students found the hint system very helpful as a learning aid, validating our design.

\begin{table}[t]
     \caption{Helpfulness ratings of Thunder CTF (1=Very Unhelpful, 2=Somewhat Unhelpful, 3=Neither Helpful nor Unhelpful, 4=Somewhat Helpful, 5=Very Helpful)}
    \begin{center}
    \begin{tabular}{|l|l|l|l|l|l|l|}
    \hline
    {\bf Question } & {\bf 1 } & {\bf 2 } & {\bf 3 } & {\bf 4 } & {\bf 5 } & {\bf Mean rating }\\
    \hline
    Q1 & 1 & 3 & 3 & 19 & 10 & 3.94\\ 
    \hline
    Q2 & 1 & 2 & 4 & 20 & 9 & 3.94\\ 
    \hline
    Q3 & 1 & 0 & 1 & 10 & 24 & 4.56\\ 
    \hline
    \end{tabular}
    \label{tab3}
    \end{center}
\end{table}

\section{Conclusion and Future Work}
\label{sec4}
With the move to cloud-based infrastructure, securing cloud applications and deployments is becoming extremely important. This paper describes a curriculum and CTF for not only teaching cloud security issues to students, but also developing their skills in applying them. Thunder CTF is designed to be applicable, extensible, and deployable, allowing it to be used in as an efficient training tool in an educational setting. Additionally, the scaffolded hint system makes the CTF accessible and informative even for novice practitioners with little background in cloud security. Results from an initial deployment of the CTF are promising, indicating that scaffolded CTFs can be effective at teaching security concepts to students. 

In future work, we will need to deploy and evaluate the use of the CTF in additional course offerings in order to attain a greater sample size. Similarly, deployments of the CTF in other course curriculums (such as security-specific courses) are necessary to understand how the technical background of students affects their experience using the CTF.

\section*{Acknowledgment}
This material is supported by the National Science Foundation under Grant No. 1821841. Any opinions, findings, and conclusions or recommendations expressed in this material are those of the author and do not necessarily reflect the views of the National Science Foundation.

\end{document}